\documentclass[review,1p,times,authoryear]{elsarticle}

\journal{Information and Software Technology}

\PassOptionsToPackage{hidelinks}{hyperref}

\usepackage[utf8]{inputenc}
\usepackage[T1]{fontenc}
\usepackage{microtype}
\usepackage{url}
\usepackage{amsmath,amssymb}
\usepackage{booktabs}
\usepackage{graphicx}
\usepackage{xcolor}
\usepackage{listings}
\usepackage{enumitem}
\usepackage{siunitx}
\usepackage{xspace}

\newcommand{\tool}{\textsc{cukereuse}\xspace}

\begin{document}

\begin{frontmatter}

\title{D\'ej\`a Vu at Scale: Paraphrase-Robust Detection of Duplicate Gherkin Steps in Behaviour-Driven Software Testing with Sentence-Transformer Embeddings and a 1.1M-Step Open Benchmark}

\author[a]{Ali Hassaan Mughal\corref{cor1}}
\ead{alihassaanmughal.work@gmail.com}
\cortext[cor1]{Corresponding author. ORCID: 0000-0002-0724-9197.}

\author[b]{Noor Fatima}
\ead{nfatima.bce25seecs@seecs.edu.pk}

\author[c]{Muhammad Bilal}
\ead{m.bilal@tum.de}

\affiliation[a]{organization={Independent Researcher; Applied MBA (Data Analytics), Texas Wesleyan University},
                city={Fort Worth, TX},
                country={USA}}

\affiliation[b]{organization={Independent Researcher; B.E.\ Computer Engineering, National University of Sciences and Technology (NUST)},
                country={Pakistan}}

\affiliation[c]{organization={Independent Researcher; M.Sc.\ Management, Technical University of Munich (ORCID: 0000-0003-4106-0256)},
                city={Munich},
                country={Germany}}

\begin{highlights}
\item Largest cross-organisational BDD step corpus released: 347 repos, 1.1M Gherkin steps
\item First public BDD-quality pair-level benchmark: 1{,}020 labels, Fleiss $\kappa = 0.84$
\item Static, paraphrase-robust step-level duplicate detector with calibrated F$_1$
\item Score-free relabelling protocol bounds rubric-detector circularity
\item ISO/IEC 25010 mapping links detected clusters to maintainability properties
\end{highlights}

\begin{abstract}
\noindent\textbf{Context.}
Behaviour-Driven Development (BDD) suites written in Gherkin
accumulate step-text duplication whose maintenance cost is
established in prior work. Existing detection techniques either
require running the tests (Binamungu et al., 2018-2023) or are
confined to a single organisation (Irshad et al., 2020-2022),
leaving a gap: a purely static, paraphrase-robust, step-level
detector usable on any repository, and a public benchmark on
which to calibrate it.

\noindent\textbf{Objective.}
We fill the gap by building (i) the largest cross-organisational
released BDD step corpus to date, (ii) a labelled pair-level
calibration benchmark for duplicate detection, and (iii) a
detection layer comprising four strategies along a
precision/compute-cost frontier, with an explicit consolidation-savings
model that maps detected clusters to ISO/IEC 25010 quality
sub-characteristics.

\noindent\textbf{Method.}
The corpus comprises 347 public GitHub repositories, 23{,}667
parsed \texttt{.feature} files, and 1{,}113{,}616 Gherkin steps,
discovered via the GitHub REST Search API and tagged with SPDX
licence identifiers. The detector layers exact hashing, normalised
Levenshtein, sentence-transformer cosine, and a hybrid
Levenshtein-banded variant. Calibration uses 1{,}020 step pairs
manually labelled by the three authors under a released rubric,
with a 60-pair overlap subset for inter-annotator agreement
(Fleiss' $\kappa = 0.84$). We report precision, recall, and F$_1$
with bootstrap 95\,\% confidence intervals under two protocols
(the primary rubric and a score-free second-pass relabelling) and
benchmark against two classical lexical baselines (SourcererCC-style
token Jaccard and NiCad-style char-$n$-gram TF-IDF).

\noindent\textbf{Results.}
The step-weighted exact-duplicate rate is 80.2\,\%; the
median-repository rate is 58.6\,\% (Spearman $\rho = 0.51$ with
size). The top hybrid cluster groups 20{,}737 occurrences across
2{,}245 files. The strongest honest pair-level number is near-exact
at F$_1 = 0.822$ on score-free labels; the primary-rubric semantic
F$_1 = 0.906$ is inflated by a stratification artefact disclosed
openly. Lexical baselines reach primary F$_1 = 0.761$ (Jaccard)
and $0.799$ (TF-IDF). The consolidation-savings model estimates
that 893{,}357 step-text occurrences across the corpus are amenable
to maintainer-action consolidation; on the median repository,
62.5\,\% of step lines are eliminable.

\noindent\textbf{Conclusion.}
A public, paraphrase-robust, step-level static detector closes a
real gap in BDD-quality tooling, and the released benchmark gives
future detector proposals a calibration target. The tool, corpus,
labelled pairs, rubric, and full analysis pipeline are released
under permissive licences at
\url{https://github.com/amughalbscs16/cukereuse-release}.
\end{abstract}


\end{frontmatter}

\section{Introduction}
\label{sec:introduction}

Gherkin, the near-English notation used by Cucumber, Behat, SpecFlow,
pytest-bdd, behave, and related Behaviour-Driven Development (BDD)
frameworks \citep{north2006bdd,wynne2017cucumber}, is deployed at
considerable scale: our survey of public GitHub places approximately
55{,}000 \texttt{.feature} files in indexable repositories, of which
this paper assembles a subset of 347 repositories
(23{,}667 parsed feature files, 1{,}113{,}616 steps;
Section~\ref{sec:corpus}). Across that corpus, more than four out of
every five steps are byte-identical duplicates of another step in
the same corpus; under paraphrase-robust matching the most-repeated
canonical step phrasing (\texttt{the response status is 200 OK})
appears 20{,}737 times across 2{,}245 files in 43 distinct
repositories.

Duplication of this shape carries documented maintenance costs.
Prior BDD-quality work establishes the cost independently:
\citet{binamungu2018vst} characterise scenario-level duplication as
a first-class quality defect; \citet{oliveira2017quality,
oliveira2019quality} catalogue practitioner-reported quality
concerns where duplication and paraphrase inconsistency recur;
\citet{irshad2021jss,irshad2022ist} study BDD reuse and the
refactoring of duplicated specifications inside a single industrial
organisation; the latter study reports automated identification of
duplicate-specification refactoring candidates roughly sixty times
faster than manual search. Three
prior bodies of work occupy the adjacent territory: Binamungu et
al.'s dynamic execution tracing
\citep{binamungu2018vst,binamungu2020phd} and accompanying
practitioner studies \citep{binamungu2018saner,binamungu2020xp}
(the tracing requires runnable tests; infeasible for cross-repository scale);
Irshad et al.'s NCD-based single-organisation static analysis
\citep{irshad2021einf,irshad2021jss,irshad2022ist} (neither neural
nor step-level); and an unpublished MSc thesis
\citep{suan2015msc}. To our knowledge, no prior work applies modern sentence
embeddings to Cucumber steps or performs duplicate detection at
step-text level across a published cross-repository corpus.

Each redundant step occurrence imposes a downstream cost: a
renamed step with 1{,}389 occurrences (the git-town
\texttt{And the branches} cluster) requires touching 1{,}389
sites unless consolidation reduces it to one. Detection enables
consolidation, paraphrase-aware lints, and corpus-grounded
reusable step libraries; the benefits map onto ISO/IEC 25010
maintainability and reliability sub-characteristics
\citep{iso25010} and supply BDD champions with a measured
artefact to anchor refactor-budget discussions in
organisations whose practitioner-reported delivery-speed/quality
trade-off is documented at scale \citep{tricentis2025qtr}.
Section~\ref{sec:benefits} reports the benefits as released-corpus
measurements without translating them into hour-saving or
cost-saving claims that the data do not support.

\subsection*{Contributions and research questions}

The paper makes six contributions: (i)~the largest released
cross-organisational BDD step corpus to date (347 repositories,
23{,}667 \texttt{.feature} files, 1{,}113{,}616 steps;
Section~\ref{sec:corpus}), released with a Gebru-style datasheet
\citep{gebru2021datasheets} and per-step Software Package Data Exchange (SPDX) licence tags;
(ii)~the first published BDD-quality labelled benchmark
(Section~\ref{sec:evaluation:labels}: 1{,}020 step pairs, three-author
labelling, Fleiss' $\kappa = 0.84$ on a 60-pair overlap
\citep{fleiss1971kappa}); (iii)~a score-free relabelling protocol
(Section~\ref{sec:evaluation:score-free}) that bounds the
circularity between score-based rubric rules and score-based
detector; (iv)~two classical lexical baselines on the same
benchmark with bootstrap 95\,\% confidence intervals (CIs)
(Section~\ref{sec:evaluation:baselines}: SourcererCC-style token
Jaccard \citep{sajnani2016sourcerercc} and NiCad-style char-$n$-gram
Term Frequency / Inverse Document Frequency, TF-IDF
\citep{cordy2011nicad,roy2008nicad}); (v)~a corpus-grounded
consolidation-savings model (Section~\ref{sec:benefits}) mapping
detected clusters to ISO/IEC 25010
\citep{iso25010} maintainability and reliability sub-characteristics;
and (vi)~the cukereuse tool itself, released as Apache-2.0 Python
alongside the corpus, benchmark, rubric, and analysis pipeline.

These contributions answer four research questions (RQs):
\textbf{RQ1} (prevalence): how prevalent is step-text duplication in
cross-organisational BDD suites and does the rate depend on licence
class or repository size? \textbf{RQ2} (detection quality): how do
four detection strategies (exact, near-exact, semantic, hybrid) and
two classical lexical baselines compare on an author-labelled
benchmark with bootstrap 95\,\% CIs? \textbf{RQ3} (methodological
coupling): how much of the reported F$_1$ reflects circularity
between score-based labels and score-based detectors? \textbf{RQ4}
(benefits and quality implications): what maintainability and
refactoring surface area is quantifiably realisable from corpus-wide
application of cukereuse, mapped to ISO/IEC 25010 dimensions?
Sections~\ref{sec:corpus}, \ref{sec:evaluation},
\ref{sec:evaluation:score-free}, and~\ref{sec:benefits} respectively
answer the four. The remainder positions the work against prior art
(Section~\ref{sec:related}, Section~\ref{sec:approach},
Section~\ref{sec:background}), enumerates threats to validity
(Section~\ref{sec:threats}), and concludes
(Section~\ref{sec:conclusion}).

\section{Background}
\label{sec:background}

\subsection{Why duplicate detection in BDD matters}
\label{sec:background:why}

Defect-discovery cost grows substantially across the software
lifecycle \citep{boehm1981software,boehm2000cocomoii}, and BDD test
suites occupy a distinctive position in this cost profile because
they serve simultaneously as acceptance specification, living
documentation, and executable test
\citep{wynne2017cucumber,smart2014bdd}. The general clone-detection
literature establishes the link between duplication,
maintainability, and refactoring outcomes:
\citet{roy2007survey} survey four clone-type classes,
\citet{krinke2008stability} reports measurable stability differences
between cloned and non-cloned code, \citet{mondal2015comparative}
report differing bug-proneness across clone types, and
\citet{alomar2025duplicateaware}
empirically links duplication-aware refactoring to specific
state-of-the-art quality metrics. The ISO/IEC 25010 product-quality
model \citep{iso25010} provides the vocabulary in which we frame
the benefit (Section~\ref{sec:benefits:iso}): step-text
consolidation directly addresses maintainability sub-characteristics
(modifiability, modularity, reusability, analyzability,
testability) and indirectly reliability via drift prevention. To
quantify how much benefit is concretely realisable from real BDD
suites at scale requires a public cross-organisational corpus, a
labelled benchmark, reported uncertainty, classical baselines, and
a model linking detected clusters to consolidation;
Sections~\ref{sec:corpus} through~\ref{sec:benefits} supply each.

\section{Related Work}
\label{sec:related}

\subsection{BDD quality and duplicate detection}

The most directly comparable line of work is due to Binamungu and
colleagues at Manchester
\citep{binamungu2018vst,binamungu2018saner,binamungu2020xp,binamungu2020phd,binamungu2023jss}.
They characterise the quality of BDD specifications, enumerate the
causes and costs of scenario duplication, and propose detection based
on recording the execution trace of each scenario and comparing the
traces for equivalence. Their technique is powerful (it can detect
behavioural equivalence of scenarios whose surface text is entirely
different) but it requires a runnable test suite in a configured
environment, a prerequisite that we avoid.

Oliveira and Marczak \citep{oliveira2017quality,oliveira2019quality}
survey and interview practitioners to catalogue the perceived quality
factors of BDD scenarios, including uniqueness (absence of duplication)
and clarity. Wautelet et al.\
\citep{wautelet2023poem} and Sears et al.\ \citep{sears2025profes}
develop quality assessment frameworks and Computer-Aided Software Engineering (CASE) tooling that analyse
individual scenarios for compliance with style guidelines. None of
these operate at scale across a published corpus of public repositories.

The work of Irshad et al.\
\citep{irshad2021einf,irshad2021jss,irshad2022ist} addresses BDD
reuse in an industrial setting within a single organisation (Ericsson)
using Normalised Compression Distance (NCD). Their focus is the
systematic reuse of acceptance tests and the refactoring of
duplicated specifications (87 specifications across two products in
the published evaluation) rather than step-level detection in
cross-organisational corpora.

Sia's 2015 Manchester Master of Science thesis \citep{suan2015msc}
prototypes a textual-plus-Abstract-Syntax-Tree (AST) Gherkin
similarity tool (SEED), unpublished outside
Manchester and evaluated on only three open-source projects
(148 feature files in total).

Table~\ref{tab:corpus-compare} positions our corpus against
comparable BDD-quality corpora in the literature, including the
concurrent GivenWhenThen (GWT) dataset
\citep{alcantara2026gwt} released at the Mining Software Repositories (MSR) 2026 conference in the same month as
this work. GWT and cukereuse share no authors and no coordination,
and their sampling strategies are deliberately different: GWT
retains on average 1.33 scenarios per repository across 1{,}720
filtered repositories (breadth-oriented sampling with
official-Cucumber-only and English-only filters), while
cukereuse exhaustively mines every \texttt{.feature} file from
each of 347 repositories (no framework filter, no language filter).
On the publicly reported axes cukereuse reports the larger
scenario count (GWT does not report a step count); on repository
count GWT reports a wider cross-section. GWT additionally links each scenario to its backing
step-definition file and source-code dependencies, a contribution
cukereuse does not make. The datasets are complementary rather
than substitutes.

\begin{table}[t]
\centering
\caption{BDD-quality corpora reported in published prior and
concurrent work, compared to the present corpus. Counts are as
reported by the cited authors; "n/r" means not reported in the
cited source. "Cross-org" is yes if the corpus spans more than
one independent organisation or repository owner. For Irshad et
al.\ the two units are industrial products of one organisation,
and the step count is derived from their reported 87
specifications at $\approx$7 steps each.}
\label{tab:corpus-compare}
\footnotesize
\setlength{\tabcolsep}{3pt}
\resizebox{\textwidth}{!}{%
\begin{tabular}{llrrrrcc}
\toprule
Corpus                                       & Year    & Repos & Files & Scenarios & Steps & Cross-org & Released \\
\midrule
Binamungu et al.\ \citep{binamungu2018vst}   & 2018    &     3 &     45 &      187 & n/r     & yes & no \\
Irshad et al.\ \citep{irshad2022ist}         & 2022    &     2 &    n/r &      n/r & $\sim$610 (internal) & no  & no \\
Sia \citep{suan2015msc}                      & 2015    &     3 &    148 &      n/r & n/r & yes & no \\
GWT \citep{alcantara2026gwt}                 & 2026    & 1{,}720 &  n/r & 2{,}289 & n/r   & yes & yes \\
\textbf{\tool\ (this paper)}                 & 2026    & \textbf{347} & \textbf{23{,}667} & \textbf{136{,}970} & \textbf{1{,}113{,}616} & \textbf{yes} & \textbf{yes} \\
\bottomrule
\end{tabular}%
}
\end{table}

A separate strand considers the quality and refactoring of
natural-language tests themselves. Soares et al.\
\citep{soares2023esem} catalogue smells in natural-language test
suites, and Aranda et al.\ \citep{aranda2024ease} catalogue
transformations that remove them, both motivated by maintainability.
Pereira et al.\
\citep{pereira2018bdd} and Scandaroli et al.\
\citep{scandaroli2019bdd} study practitioners' experience with BDD
adoption, and Nascimento et al.\ \citep{nascimento2020bdd} evaluate
its benefits and challenges through an expert panel. None surface
duplicate detection at corpus scale.

\section{Approach}
\label{sec:approach}

cukereuse layers four detection strategies along a precision /
compute-cost frontier: exact (BLAKE2b hashing on whitespace-collapsed
text), near-exact (Levenshtein ratio), semantic (Sentence-BERT
cosine \citep{reimers2019sbert,wang2020minilm}, abbreviated SBERT
hereafter), and hybrid (SBERT cosine combined with a Levenshtein-band
guard). Cheap strategies run first so that expensive strategies only
adjudicate the pairs the cheap strategies could not resolve.

\subsection{Identity definition}
\label{sec:approach:identity}

For the purposes of duplicate detection, the \textbf{identity} of a
step is defined as
\[
\text{id}(s) = \mathrm{BLAKE2b}\bigl(\mathrm{whitespace\_collapse}(s.\text{text})\bigr)
\]
where $s.\text{text}$ is the step phrasing line minus the keyword
(\texttt{Given}/\texttt{When}/\texttt{Then}/\texttt{And}/\texttt{But})
and minus any attached DocString or DataTable argument. Quoted
parameters (\texttt{"admin"}), Scenario-Outline placeholders
(\texttt{<role>}), numeric literals (\texttt{5}), URLs, and
identifiers are part of the identity. This definition mirrors how
the Cucumber runtime resolves a step against its step-definition
function: the runtime ignores the leading keyword, hands the
step-phrasing line to a regular-expression match against
step-definition signatures, and passes any attached DocString or
DataTable to the matched function as a separate argument.

The exclusion of DocString and DataTable bodies is both
semantically motivated (the Gherkin grammar specifies them as
\emph{step arguments}, not as part of the step phrasing,
\citealp{gherkin2023}) and technically necessary: multi-line
JSON / SQL / HTML payloads exceed the 256-token input window of
the all-MiniLM-L6-v2 sentence-transformer embedder (a distillation
in the MiniLM family, \citealp{wang2020minilm}), would
fragment exact-duplicate hashing such that the headline 80.2\,\%
rate would collapse to single-digit percentages, and would
inflate Levenshtein computation cost by two to three orders of magnitude.
Section~\ref{sec:threats:construct} discusses the construct-validity
implications of this choice.

\section{Corpus Construction}
\label{sec:corpus}

\begin{figure}[t]
\centering
\includegraphics[width=\textwidth]{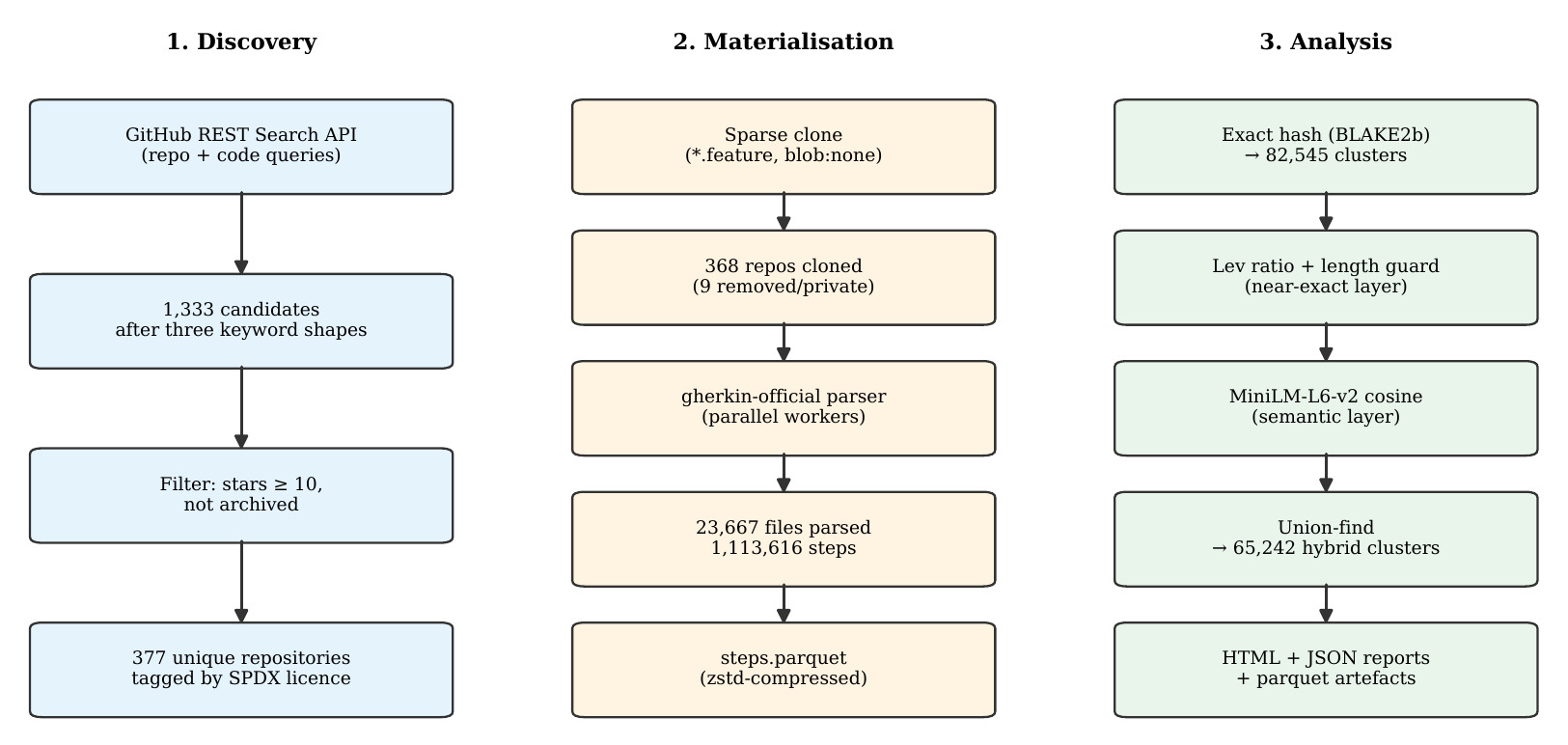}
\caption{End-to-end pipeline for cukereuse corpus construction
and analysis. Discovery surfaces 377 unique repositories via the
GitHub REST Search API after deduplication and the
$\geq 10$-stars / not-archived filters. Materialisation sparse-clones
the \texttt{.feature}-file blobs of those repositories (368
succeed; 9 had been removed or privatised in the interval between
discovery and clone) and parses them with the official Gherkin
parser, of which 347 yield at least one parseable step and form
the released corpus. Analysis layers the four detection strategies
(BLAKE2b $\rightarrow$ Levenshtein $\rightarrow$ MiniLM-L6-v2
$\rightarrow$ union-find) and emits browsable HTML, machine-readable
JSON, and columnar parquet artefacts.}
\label{fig:pipeline}
\end{figure}

The corpus-construction pipeline is summarised in
Figure~\ref{fig:pipeline}. Three stages (discovery,
materialisation, analysis) run end-to-end in approximately three
hours of wall-clock time on a commodity workstation; each stage
caches its artefacts so that downstream stages can re-run without
repeating earlier work.

\subsection{Discovery}
\label{sec:corpus:discovery}

Discovery uses the GitHub REST Search API
\citep{github2024restsearch} with two complementary query shapes:
\texttt{/search/repositories} filtered by Linguist
\citep{linguist2024} primary language Gherkin (yielding $\approx 171$
candidates at $\geq 10$ stars, not-archived) and \texttt{/search/code}
with three keyword-plus-extension filters
(\texttt{"Feature:"}, \texttt{"Scenario:"}, \texttt{"Background:"} on
\texttt{.feature}; yielding $\approx 1{,}162$ additional candidates).
The two shapes are complementary because \texttt{language:Gherkin}
captures only Gherkin-dominant repositories whereas the much larger
population of BDD use occurs in projects where Gherkin is a
minority language; three keyword variants hedge against the
1{,}000-result-per-query REST cap. After deduplication by
\texttt{owner/name} and re-application of the $\geq 10$-stars and
not-archived filters, 377 unique repositories survive. The stars
threshold is a standard MSR-sample hygiene decision (e.g.,
\citealp{munaiah2017sample,borges2016github,kalliamvakou2014ghq}):
$\geq 1$ admits tutorial clones that inflate duplication;
$\geq 100$ would collapse the sample to 36 repositories.

\subsection{Corpus headline figures}

\begin{itemize}[leftmargin=2em]
\item 347 repositories, 23{,}667 parsed \texttt{.feature} files, 1{,}113{,}616
      steps, 220{,}312 unique normalised step texts.
\item 61{,}214 background steps (5.5\,\%) and 66{,}020 outline steps
      (5.9\,\%).
\item Exact-duplication rate: 80.22\,\%.
\item 82{,}545 clusters under exact detection; 65{,}242 under hybrid.
\item Top cluster under exact: \texttt{the request is sent}
      (16{,}815 occurrences).
\item Top cluster under hybrid: \texttt{the response status is 200 OK}
      (20{,}737 occurrences across 2{,}245 distinct files).
\item License mix at step level:
      permissive 635{,}586 (57.1\,\%);
      copyleft 232{,}297 (20.9\,\%);
      unknown 174{,}077 (15.6\,\%);
      unlicensed 71{,}656 (6.4\,\%).
\end{itemize}

The step-text length distribution (over all 1{,}113{,}616 step
occurrences) is right-skewed: the median step is 42 characters,
the p25/p75 inter-quartile range is [29, 62] characters, and the
p99 is 162 characters (long parametric step texts with substantial
inline arguments). Over the 220{,}312 unique normalised step texts
the distribution shifts modestly upward (median 51, p99 197), as
short-and-frequently-repeated phrasings deduplicate but long-and-
unique ones do not.

\subsection{Why this dataset matters}
\label{sec:corpus:why}

Three properties of the released artefact place it in a different
category from prior BDD-quality data. \emph{First, scale:}
1{,}113{,}616 steps across 23{,}667 \texttt{.feature} files exceed
the concurrent GWT dataset by a factor of roughly sixty in scenario
count and prior BDD-quality evaluation corpora by more than three
orders of magnitude in step count
(Table~\ref{tab:corpus-compare}); prior work
of \citet{binamungu2018vst,irshad2022ist,suan2015msc} relies on
corpora of at most three projects each --- 187 scenarios,
approximately six hundred industrial steps, and 148 feature files,
respectively. \emph{Second,
cross-organisational coverage:} the corpus spans 347 repositories
from 284 distinct GitHub owner accounts, spanning multiple BDD frameworks
(Cucumber-JVM, behave, pytest-bdd, Cucumber-Ruby, cucumber-js,
SpecFlow, Behat, Karate, and minor variants), supporting findings
about BDD practice that no single-organisation study can
substantiate. \emph{Third, release engineering for reproducibility
under licence diversity:} each parquet row carries the upstream
repository's SPDX licence identifier; raw \texttt{.feature}-file
bodies are not redistributed (many inherit copyleft obligations)
but are reconstructible on demand from pinned commit SHAs by the
released \texttt{rehydrate.py} script. The accompanying Gebru-style
datasheet \citep{gebru2021datasheets} documents collection,
filtering, biases, and intended uses. The 1{,}020-pair labelled
calibration benchmark with three-author Fleiss' $\kappa = 0.84$
inter-annotator agreement is, to our knowledge, the first
published BDD-quality labelled benchmark; future detector
proposals can calibrate against it directly.

\section{Evaluation}
\label{sec:evaluation}

\subsection{Labelled pairs}
\label{sec:evaluation:labels}

We sample 1{,}020 step-text pairs stratified in 170-pair bins across
six cosine-similarity bands ($[0.50, 0.70)$, $[0.70, 0.80)$,
$[0.80, 0.85)$, $[0.85, 0.90)$, $[0.90, 0.95)$, $[0.95, 1.00)$),
chosen to give adequate coverage of the threshold-sensitivity
region rather than to be representative of the underlying corpus
(which is overwhelmingly negative). Three authors label the pairs
under a shared written rubric (released as
\texttt{corpus/LABELING\_RUBRIC.md}); the rubric defines ten
ordered decision rules R1-R10 (parametric variants, synonym
paraphrase, and structural swap as positive rules R1-R3;
framework-keyword, HTTP-verb, polarity, presence-vs-content, and
action-vs-assertion distinctions as negative rules R4-R8;
outline-placeholder rename as positive R9; default-not-duplicate as
R10). Mughal labelled 500 pairs, Fatima 300, Bilal 220.

A separate 60-pair stratified overlap subset is labelled
independently by all three authors before the main labelling
batch. On the overlap, raw agreement is 53 of 60 pairs (88.3\,\%)
and Fleiss' $\kappa = 0.84$ \citep{fleiss1971kappa,landis1977kappa};
the overlap labels and disagreement analysis are released as
\texttt{corpus/labeled\_pairs\_overlap.jsonl}. Every row of
\texttt{labeled\_pairs.jsonl} records the rule that fired so that
rubric application is auditable pair-by-pair without
re-annotation. Of the 1{,}020 pairs, 494 are positive and 526
negative (48.4\,\% positive rate). Section~\ref{sec:evaluation:score-free}
introduces a score-free second-pass relabelling that bounds the
circularity introduced by the score-based positive rules R1-R3.

\subsection{Lexical baselines}
\label{sec:evaluation:baselines}

The clone-detection literature has mature token- and lexical-signature-based
detectors that have been evaluated extensively on source code but not,
to our knowledge, on BDD step text. To position the cukereuse numbers
against this baseline, we run two lexical strategies on the same
1{,}020-pair benchmark, following the defining primitive of two
well-known detectors:

\begin{description}
\item[Token-set Jaccard (SourcererCC-style).] We aggressively normalise
each step text, replacing quoted parameters with a \texttt{PARAM}
placeholder, lowercasing, stripping punctuation, collapsing
whitespace, then compute the Jaccard similarity of the two token
sets. SourcererCC \citep{sajnani2016sourcerercc} uses the same
token-multiset-overlap primitive at the block and method
level on source
code; we add text normalisation and adapt it to pair-of-strings
at the sentence level.
\item[TF-IDF character $n$-gram cosine (NiCad-style).] We fit a
character-3-to-5-gram TF-IDF vectoriser on all step texts in the
labelled set, then compute the cosine of the TF-IDF vectors of the
two texts in each pair. NiCad \citep{cordy2011nicad} uses a normalise-
plus-pretty-print-plus-compare primitive; our character-$n$-gram
cosine captures the lexical-signature spirit of that primitive
without requiring NiCad's code-aware pretty-printer.
\end{description}

Both baselines are implemented in \texttt{scripts/revision\_analyses.py}
for reproducibility. Table~\ref{tab:calibration} includes their
best-F$_1$ operating points alongside the cukereuse strategies.

\subsection{Score-free second-pass relabelling}
\label{sec:evaluation:score-free}

The most serious threat to the primary calibration is circularity:
rules R1, R2, and R3 of the rubric use cosine and Levenshtein
cut-points, so the detector's best-F$_1$ threshold partly recovers
the rubric's own cut-points rather than measuring agreement with an
independent reference. To bound this circularity, we perform a
second-pass relabelling of the entire 1{,}020-pair set under a
protocol that never accesses cosine or Levenshtein. The score-free
protocol reuses the structural rules R4-R8 of the primary rubric
and replaces the score-based positive rules R1-R3 with four
deterministic text-rewriting rules: parametric-normalised identity
plus hand-curated BDD synonym canonicalisation; token-multiset
identity after canonicalisation; subsequence containment of at
least 70\,\% of the shorter text; and token-set Jaccard at least
0.80. None of these rules consults a learned similarity score; the
protocol is implemented verbatim in
\texttt{scripts/revision\_analyses.py}.

The relabelling assigns 565 positives and 455 negatives (55.4\,\%
positive rate vs the primary rubric's 48.4\,\%); Cohen's $\kappa$
between protocols is 0.470 ("moderate"), with raw disagreement
on 271 of the 1{,}020 pairs (26.6\,\%, against an
expected-by-chance disagreement of 50.2\,\%). The honest pair-level F$_1$ under
score-free labels is the headline number reported in
Table~\ref{tab:calibration}.

Table~\ref{tab:calibration} is our headline calibration table. It
reports for each strategy: the best-F$_1$ threshold under the primary
rubric, the corresponding precision and recall with bootstrap
95\,\% confidence intervals, and F$_1$ against the score-free
relabelling. The two lexical baselines are included for direct
comparison.

\begin{table}[t]
\centering
\caption{Best-F$_1$ operating point per strategy on the 1{,}020-pair
labelled set. \emph{Primary} columns evaluate against the primary
rubric (494 positives, 526 negatives) with bootstrap 95\,\%
confidence intervals in brackets; the \emph{Score-free} column
evaluates against the score-free relabelling (565 positives) of
Section~\ref{sec:evaluation:score-free}. Bold marks the strongest
pair-level number (near-exact, F$_1 = 0.822$ score-free); semantic's
primary-rubric F$_1 = 0.906$ is shown for completeness but reflects
the specific rubric cut-points.}
\label{tab:calibration}
\footnotesize
\setlength{\tabcolsep}{3pt}
\resizebox{\textwidth}{!}{%
\begin{tabular}{llcccc}
\toprule
Strategy                               & Thr.  & P (primary, 95\,\% CI) & R (primary, 95\,\% CI) & F$_1$ primary                    & F$_1$ score-free \\
\midrule
semantic (SBERT)                        & 0.82  & 0.828 [0.796, 0.856]   & 1.000\textsuperscript{$\ast$} [1.000, 1.000]   & 0.906 [0.886, 0.923]    & 0.772 \\
near-exact (Lev)                        & 0.80  & 0.827 [0.797, 0.858]   & 0.901 [0.875, 0.926]   & 0.862 [0.841, 0.885]             & \textbf{0.822} \\
hybrid (Lev-band $[0.3, 0.95]$)         & 0.82  & 0.772 [0.731, 0.811]   & 0.680 [0.638, 0.719]   & 0.723 [0.690, 0.753]             & 0.625 \\
\midrule
\emph{baseline:} token-set Jaccard      & 0.56\textsuperscript{$\dagger$} & 0.636 & 0.947 & 0.761 & 0.979\textsuperscript{$\dagger\dagger$} \\
\emph{baseline:} TF-IDF char $n$-gram   & 0.50\textsuperscript{$\dagger$} & 0.717 & 0.903 & 0.799 & 0.755\textsuperscript{$\dagger\dagger$} \\
\bottomrule
\end{tabular}%
}

\vspace{0.3em}
\footnotesize\noindent\textsuperscript{$\ast$} Recall = 1.000 is a stratification artefact of the label
set, not a detector property: every primary-rubric positive has
cosine $\geq 0.80$ by rule R1-R3, and the detector's threshold of
0.82 lies within the same band. Precision and F$_1$ are the
discriminating numbers.
\textsuperscript{$\dagger$} Baseline thresholds in the Thr. column are
tuned to the primary rubric; each baseline was separately tuned to
the score-free labels (best threshold 0.80 for Jaccard, 0.34 for
TF-IDF) for the score-free F$_1$ column.
\textsuperscript{$\dagger\dagger$} Token-Jaccard's F$_1 = 0.979$
against score-free labels reflects the score-free protocol's
token-overlap positive rules (P1-P4); see
Section~\ref{sec:evaluation:score-free}.
\end{table}

Three observations on Table~\ref{tab:calibration}. \emph{First,}
semantic's recall of 1.000 under the primary rubric is a
labelling-stratification artefact (rules R1-R3 require cosine
$\geq 0.80$ and the detector threshold sits inside that band),
not an independent detector property; the discriminating numbers
are precision and the score-free F$_1$. \emph{Second,} the strategy
ranking flips between protocols: near-exact wins on score-free
labels (F$_1 = 0.822$ versus semantic's 0.772) while semantic wins
under the primary rubric. Near-exact's primary-to-score-free drop
of 0.04 is also the smallest of the four detector strategies, making
it the most cross-protocol-stable pair classifier on this benchmark.
\emph{Third,} all three cukereuse strategies beat the lexical
baselines under the primary rubric by 0.07 to 0.15 F$_1$; under
score-free labels the Jaccard baseline reaches F$_1 = 0.979$ because
the score-free positive rules P1-P4 are themselves token-based, so
a token-set classifier recovers them by construction. The
score-free protocol is \emph{score-independent}, not feature-
independent.

\subsection{Held-out cross-validation}
\label{sec:evaluation:cv}

The thresholds in Table~\ref{tab:calibration} are tuned on the same
1{,}020-pair set on which F$_1$ is reported, so the headline numbers
are in-sample. To bound the resulting optimism we additionally run
5-fold stratified cross-validation: each fold's threshold is selected
on the training partition and F$_1$ is evaluated on the held-out
partition. The held-out CV F$_1$ values (mean $\pm$ standard
deviation across folds) are: near-exact $0.863 \pm 0.026$
(range $[0.835, 0.895]$), semantic $0.906 \pm 0.007$
(range $[0.895, 0.912]$), hybrid $0.723 \pm 0.035$
(range $[0.685, 0.756]$). Each held-out mean is within $0.001$ of
the in-sample headline F$_1$ in Table~\ref{tab:calibration}, so
threshold selection on this benchmark contributes negligible
optimism. The CV procedure is implemented in
\texttt{scripts/cv\_calibration.py}; its full per-fold output is
in \texttt{analysis/cv\_calibration.json}.

\section{Benefits and Software Quality Implications}
\label{sec:benefits}

This section answers RQ4: what maintenance, refactoring,
readability, drift-risk-reduction, and reusable-library benefits
are quantifiably realisable from corpus-wide application of the
detection layer described in
Sections~\ref{sec:approach} through~\ref{sec:evaluation}? The answer is
quantitative where possible and qualitative where appropriate, and
is grounded throughout in observations from the released corpus
and the benchmark evaluation.

\subsection{ISO/IEC 25010 quality-dimension mapping}
\label{sec:benefits:iso}

The ISO/IEC 25010 product-quality model \citep{iso25010} defines
eight characteristics of product quality. Five sub-characteristics
of \emph{maintainability} are directly addressed by step-text
duplicate detection (Table~\ref{tab:iso25010-mapping}); a sixth
characteristic, \emph{reliability}, is addressed indirectly via
drift prevention. The remaining ISO~25010 characteristics
(functional suitability, performance efficiency, compatibility,
usability of the system under test, security, portability) are
out of scope for the present work.

\begin{table}[t]
\centering
\caption{Mapping of cukereuse contributions to ISO/IEC 25010
\citep{iso25010} quality characteristics.}
\label{tab:iso25010-mapping}
\footnotesize
\setlength{\tabcolsep}{4pt}
\resizebox{\textwidth}{!}{%
\begin{tabular}{p{3.0cm}p{2.6cm}p{6.5cm}p{1.6cm}}
\toprule
ISO~25010 char. & Sub-char. & How cukereuse addresses it & Section \\
\midrule
Maintainability & Modifiability    & Cluster-level rename: $O(1)$ per cluster vs $O(N)$ per occurrence; the $N{=}1{,}389$ case in git-town becomes a single edit. & §\ref{sec:benefits:savings} \\
Maintainability & Modularity        & Top hybrid clusters surface cohesive step vocabulary candidates for shared-library extraction. & §\ref{sec:benefits} \\
Maintainability & Reusability       & 65{,}242 hybrid clusters with cross-organisational distribution; the top-20 cover ${\sim}10\,\%$ of step occurrences. & §\ref{sec:benefits} \\
Maintainability & Analyzability     & Cluster reports surface implicit suite structure that scenario-by-scenario reading misses. & §\ref{sec:benefits:savings} \\
Maintainability & Testability       & Released benchmark with Fleiss' $\kappa = 0.84$ enables future detector comparison. & §\ref{sec:evaluation} \\
Reliability    & Maturity          & Paraphrase-aware lints prevent silent drift between phrasing and glue-code function. & §\ref{sec:benefits} \\
\bottomrule
\end{tabular}%
}
\end{table}

\subsection{Consolidation-savings model}
\label{sec:benefits:savings}

\begin{figure}[t]
\centering
\includegraphics[width=0.78\textwidth]{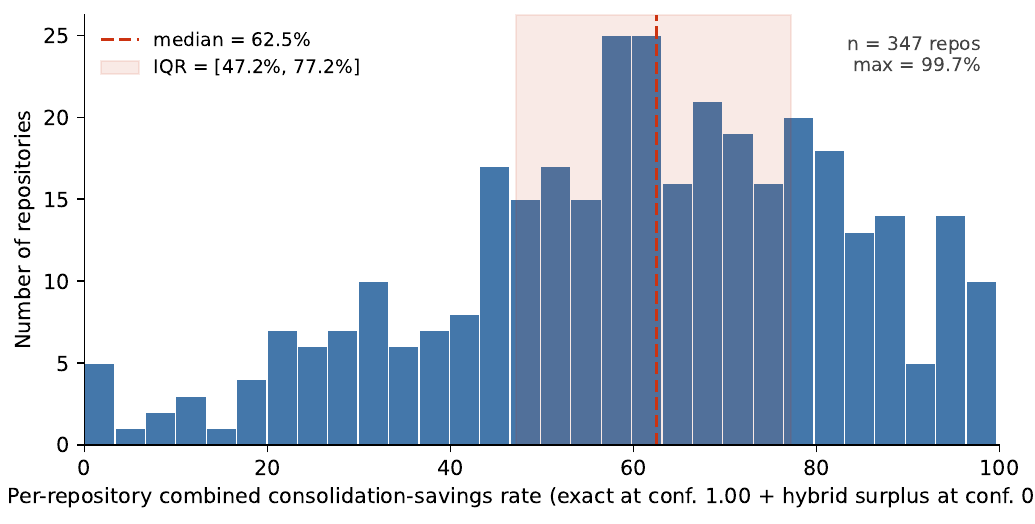}
\caption{Per-repository consolidation-savings rate across the 347
corpus repositories. Combined rate = exact at confidence 1.00 +
hybrid surplus over exact at confidence 0.57 (the strategy's
score-free precision on the labelled benchmark;
Equation~\ref{eq:savings-cluster}). Median 62.5\,\%, IQR
[47.2\,\%, 77.2\,\%], maximum 99.7\,\%.}
\label{fig:savings-hist}
\end{figure}

For a cluster $c$ produced by detection strategy $s$ with
$\text{occurrences}(c)$ members and a strategy-specific confidence
factor $\text{conf}(s) \in [0,1]$, define the consolidatable
occurrences of $c$ as
\begin{equation}
\text{savings}(c) = (\text{occurrences}(c) - 1) \cdot \text{conf}(s)
\label{eq:savings-cluster}
\end{equation}
i.e., one canonical occurrence is kept and the remainder are
candidates for elimination, scaled by an explicit confidence
factor. Each $\text{conf}(s)$ is anchored in the corresponding
strategy's empirically measured precision on the 1{,}020-pair
labelled benchmark of Section~\ref{sec:evaluation}: we use
$\text{conf}(\text{exact}) = 1.00$ (byte-identical and therefore
definitionally certain), $\text{conf}(\text{near-exact}) = 0.83$
(matching the strategy's primary-rubric precision $0.827$), and
$\text{conf}(\text{hybrid}) = 0.57$ (matching the strategy's
\emph{score-free} precision $0.569$, deliberately the more
conservative of the two protocols for the noisier strategy).
The per-repository savings rate is
$\text{savings}_r / n_{\text{steps},r}$ summed over clusters
attributed to $r$. Across the 347-repository corpus the rate
distribution (Figure~\ref{fig:savings-hist}) has median
\textbf{62.5\,\%}, inter-quartile range (IQR) [47.2\,\%, 77.2\,\%], maximum 99.7\,\%; 247
of 347 repositories (71\,\%) exceed 50\,\%. The corpus-wide
aggregate under the conservative model (exact-only at full
confidence) is \textbf{893{,}357 step-text occurrences},
identical to the 80.2\,\% headline of Section~\ref{sec:corpus}, now
interpreted as the maximum eliminable surface. Including the
hybrid-surplus over exact at the empirically anchored $0.57$ raises
the aggregate to \textbf{934{,}884} occurrences; sensitivity to the
hybrid confidence factor across $[0.0, 1.0]$ bounds the aggregate
between 893{,}357 and 966{,}212. We do not translate eliminable lines to
maintainer-hours or monetary savings: that translation requires
per-organisation cost calibration which the released
\texttt{cukereuse} CLI exposes via per-cluster member rosters but
which we cannot ground in published per-line maintenance-hour
benchmarks specific to BDD test code.

\subsection{Industry context for the proposed solution}
\label{sec:benefits:industry-need}

Two reported quantitative findings from the 2025 Tricentis Quality
Transformation Report \citep{tricentis2025qtr} (a survey of 2{,}750
software-delivery practitioners across ten countries and five
industry verticals) frame the practitioner-side context:
45\,\% of organisations report prioritising delivery speed over
software quality (against 13\,\% prioritising quality), and 42\,\%
report that poor software quality costs their organisation
\$1M or more annually. These are reported survey findings, not
quality-cost ground truth. The classic software-engineering
literature establishes that defect-discovery cost grows
substantially across the development lifecycle
\citep{boehm1981software,boehm2000cocomoii}, and BDD test suites
occupy a distinctive position because they serve simultaneously as
acceptance specification, living documentation, and executable test
\citep{wynne2017cucumber,smart2014bdd}.

No published tooling or empirical artefact currently fills the
specific gap that cukereuse addresses: a paraphrase-robust,
static, cross-framework, cross-organisational duplicate-step
detector with a publicly-released labelled benchmark. Existing
dynamic techniques \citep{binamungu2018vst} require executable
test infrastructure; existing single-organisation static
techniques \citep{irshad2022ist} have not been validated
cross-organisationally or released; classical clone detectors
\citep{sajnani2016sourcerercc,cordy2011nicad} target
programming-language source rather than near-natural-language
step text. The position of the present work is not that this
gap is the dominant determinant of BDD-suite quality (we
have no causal evidence to that effect) but that the gap is
specific, measurable, and now filled by released artefacts that
allow others to replicate and extend the analysis.

\subsection{How the eliminable-lines surface scales with suite complexity}
\label{sec:benefits:savings-tiers}

To characterise how the consolidation-savings surface distributes
across project sizes, we group the 347 corpus repositories into
four step-count tiers and report two facts per tier: how many
repositories sit in the tier, and the total eliminable-line
surface in the tier under the conservative consolidation model.
Both quantities are direct measurements over the released corpus
(Table~\ref{tab:savings-tiers}). We \emph{do not} translate
eliminable lines to person-hours or to monetary savings: such a
translation would require per-organisation maintenance-cost
calibration which is out of scope here and which we cannot ground
in published per-line maintenance-hour benchmarks for BDD test
code specifically. The released \texttt{cukereuse} CLI exposes
the per-repository eliminable-line count and the per-cluster
member roster, leaving the cost translation to the maintainer who
knows their own team's per-line maintenance burden.

\begin{table}[t]
\centering
\caption{Distribution of eliminable-line surface across repository-size
tiers. Counts are direct measurements: the repository step counts come
from \texttt{steps.parquet}; the eliminable-line surface is the sum of
the conservative consolidation-savings model
(Equation~\ref{eq:savings-cluster}, exact-match clusters at confidence
1.00) over repositories in each tier.}
\label{tab:savings-tiers}
\footnotesize
\setlength{\tabcolsep}{4pt}
\begin{tabular}{p{2.4cm}p{2.5cm}r r r}
\toprule
Tier & Step count & Repos in corpus & Total tier steps & Tier eliminable lines \\
\midrule
Small      & ${<}\,1{,}000$        & 240 & 64{,}181  & 38{,}742  \\
Medium     & 1{,}000 to 10{,}000   & 82  & 247{,}828 & 175{,}706 \\
Large      & 10{,}000 to 100{,}000 & 24  & 624{,}000 & 504{,}455 \\
Enterprise & ${\geq}\,100{,}000$   & 1   & 177{,}607 & 174{,}453 \\
\bottomrule
\end{tabular}
\end{table}

The distribution is highly concentrated. The Large tier (24
repositories, 6.9\,\% of corpus by repository count) accounts for
56.0\,\% of total corpus steps; the single Enterprise-tier
repository accounts for an additional 16.0\,\%; the Small tier
accounts for 5.8\,\% of total steps despite holding 69.2\,\% of
corpus repositories. The eliminable-line surface inherits this
concentration: any team working on a Large or Enterprise BDD
suite will encounter substantially more consolidation candidates
than a team on a Small suite. We claim no specific
maintenance-hour or monetary saving from these line counts; the
conversion depends on the maintainer's local cost model.

\subsection{Three illustrative cluster patterns}
\label{sec:benefits:cases}

Three clusters from the released report illustrate the range of
patterns the corpus surfaces. We report only the released-data
facts; we do not claim what the original authors intended or
prescribe specific refactorings.

\textbf{Within-repository concentration.} The cluster
\texttt{I use an authentication token} appears 5{,}880 times in 213
of the 218 \texttt{.feature} files of one repository
(\texttt{keygen-sh/keygen-api}, total step count 71{,}608 across
6{,}020 scenarios). The phrasing is part of the scenario template
used by approximately 88\,\% of the repository's scenarios.

\textbf{DataTable-led intro pattern.} The cluster \texttt{the
branches} appears 1{,}389 times across 1{,}187 files of one
repository (\texttt{git-town/git-town}); each occurrence is
followed by a DataTable whose contents differ. cukereuse's
identity rule (Section~\ref{sec:approach:identity}) excludes the
attached DataTable from the cluster key, so the 1{,}389 occurrences
group as one cluster; whether to consolidate is a maintainer
decision the tool does not make.

\textbf{Cross-repository convergence.} The cluster \texttt{I add
"Accept" header equal to "application/json"} appears 3{,}589 times
across 1{,}004 files in 21 distinct repositories owned by 11
distinct GitHub accounts. The corpus surfaces an HTTP-testing
phrasing on which 11 unrelated organisations have independently
converged.

\section{Threats to Validity}
\label{sec:threats}

\subsection{Internal validity: rubric-detector coupling}

The 1{,}020 labels used for the calibration study were produced by
the three authors under the shared written rubric (500 pairs by
Mughal, 300 by Fatima, 220 by Bilal), with the per-pair rule record
released in \texttt{labeled\_pairs.jsonl}
(Section~\ref{sec:evaluation:labels}). Inter-annotator Fleiss'
$\kappa$ on the 60-pair overlap subset is 0.84 (almost perfect
agreement on the Landis-Koch scale), so the labels themselves are
consistent across annotators. The residual methodological question is the rubric's
coupling with the detector's features. Rules R1-R3 of the primary rubric
use cosine and Levenshtein cut-points as labelling criteria, so
primary-rubric F$_1$ partly measures agreement between the detector
and the rubric's own cut-points. We quantify this coupling directly
via the score-free second-pass relabelling of
Section~\ref{sec:evaluation:score-free}: Cohen's $\kappa$ between
primary and score-free labellings is 0.47, detector F$_1$ drops by
0.04-0.13 depending on strategy (near-exact is the most stable at
$\Delta F_1 = 0.04$), and the token-set Jaccard baseline reaches
F$_1 = 0.979$ against score-free labels (a complementary coupling
signature of that protocol). Rubric choice therefore leaves a
signature on the evaluation; we report both rubric-vs-detector and
rubric-vs-baseline numbers so readers can judge each protocol's
coupling independently.

\subsection{External validity: sample bias and size confound}

Our 347-repository corpus is biased toward repositories that (i) have
Gherkin files discoverable by GitHub's Code Search, (ii) have
$\geq 10$ stars, (iii) are not archived. This biases against very
small, private, archived, or single-author projects. The discovery
also misses repositories whose feature-file paths do not match our
search heuristics. The BigQuery path
\citep{ghtorrent2012,gharchive} would increase coverage and is a
concrete direction for follow-up work. Our sample is nonetheless,
to our knowledge, an order of magnitude larger than any prior
cross-repository BDD corpus.

A second external-validity consideration is the repository-size
confound documented in Section~\ref{sec:corpus}: per-repository
duplication rate is strongly correlated with repository size
(Spearman $\rho = 0.51$). The pooled 80.2\,\% headline is a step-weighted
statistic, dominated by a minority of very large repositories; the
median repository's intra-repository duplication rate is only
58.6\,\%. Readers should compare their own codebases against the
median, not the pooled rate. Licence-class differences are also
partly attributable to size: the unlicensed stratum's mean
repository step count is roughly 1{,}100 against approximately
3{,}000 for the permissive stratum.

\subsection{Construct validity: what counts as a duplicate}
\label{sec:threats:construct}

Our operational definition of ``duplicate'' is \emph{would a
maintainer consolidate these two step phrasings into one step
definition?} This is deliberately liberal (it catches parametric
variants) and is not aligned with the narrower definitions used in
classical clone-detection literature
\citep{roy2007survey,bellon2007comparison}. A project that
deliberately maintains two step definitions for similar phrasings
with divergent glue code would flag as a false positive under our
definition. The maintainer-intent definition is appropriate for the
intended use case (consolidation suggestions for BDD refactoring),
but the mismatch with the classical clone-detection taxonomy should
be noted by readers comparing across literatures.

\paragraph{Mapping to Roy-Cordy clone-type categories.}
To bridge to the classical taxonomy \citep{roy2007survey}, the four
cukereuse strategies map as follows. \emph{Exact} (BLAKE2b on
whitespace-collapsed text) corresponds to \textbf{Type~1}
(identical fragments up to whitespace, layout, and comment
variation); we keep parameters and literals as part of the
identity, since renaming-tolerant matching belongs to Type~2 in
the canonical taxonomy. We measure 82{,}545
such clusters covering 975{,}902 step occurrences. \emph{Near-exact}
(Levenshtein ratio $\geq 0.80$) targets \textbf{Type~2}
(parameterised clones with small lexical variations). \emph{Semantic}
(SBERT cosine $\geq 0.82$ alone) reaches into \textbf{Type~3}
(near-miss with lexical divergence) and \textbf{Type~4} (semantically
equivalent paraphrase). \emph{Hybrid} (semantic cosine plus a
Levenshtein band of $[0.3, 0.95]$) is deliberately scoped to
Type~2/3 territory: the lower bound $0.3$ excludes pure Type~4
clones (paraphrases with no token overlap), and the upper bound
$0.95$ excludes near-Type~1 cases already captured by exact
detection. Beyond the 975{,}902 Type~1 occurrences captured by
exact detection, hybrid absorbs an additional 55{,}552 occurrences
of Type~2/3 clones into 65{,}242 unioned clusters. Pure Type~4
clones (low token overlap, high semantic similarity) are
\emph{not} a target of cukereuse's hybrid layer; readers comparing
against Type-4-aware detectors should note this scope decision.

\subsection{Conclusion validity: parameter sensitivity}

The decision thresholds reported in Section~\ref{sec:evaluation} are
empirical knees on our corpus. On different corpora with different
vocabulary profiles (e.g., a BDD suite dominated by non-English
locale) the optimal thresholds will differ. Re-running calibration
against a small project-local labelled set is the safe practice
before adopting \tool's default thresholds for a specific codebase;
the \texttt{cukereuse calibrate} CLI command performs this run.

\section{Conclusion and Future Work}
\label{sec:conclusion}

\subsection*{Conclusion}

This paper presents \tool, the first static, paraphrase-robust,
cross-organisational duplicate-step detector for Gherkin, alongside
a 1{,}113{,}616-step public corpus (347 repositories), a 1{,}020-pair
labelled benchmark with three-author Fleiss' $\kappa = 0.84$, and a
score-free relabelling protocol that bounds rubric-detector
circularity. The honest pair-level F$_1 = 0.822$ achieved by the
near-exact strategy on score-free labels stands as the calibration
target on this benchmark; classical lexical baselines
(SourcererCC-style and NiCad-style) reach 0.761 and 0.799
respectively. The corpus-grounded consolidation-savings model
estimates 893{,}357 step-text occurrences are amenable to
maintainer-action consolidation under exact-match merges, with a
median per-repository eliminable rate of 62.5\,\% (247 of 347
repositories above 50\,\%). We do not translate eliminable lines
to monetary or person-time savings; that translation requires
per-organisation maintenance-cost calibration which the released
\texttt{cukereuse} CLI exposes via per-repository cluster
membership rosters. The release is open source; the empirical
artefacts
(corpus, benchmark, rubric, datasheet, pipeline) are first-of-kind
contributions to the BDD-quality literature and are immediately
reusable by future detector proposals.

\subsection*{Future Work}
\label{sec:future}

Direct extensions of the released primitives include tooling
(project-wide find-usages, paraphrase-aware pre-commit lints,
Scenario-Outline refactoring assistants); glue-code analysis (step-
definition bodies; addresses the silent-drift failure mode and is
tractable across the corpus's primary languages: Python,
JavaScript, Ruby, Java); a larger corpus via a BigQuery discovery
path lifting the REST Search 1{,}000-result cap; running
reference-implementation lexical baselines (CCFinder, DECKARD,
NiCad, SourcererCC binaries) against our corpus; model ablation at
larger sentence-embedding sizes; and calibration on non-English
Gherkin suites whose localised keywords our English-only corpus
does not exercise.

\section*{CRediT authorship contribution statement}

\noindent\textbf{Ali Hassaan Mughal:} Conceptualization, Data
curation, Formal analysis, Investigation, Methodology, Project
administration, Software, Validation, Visualization, Writing --
original draft, Writing -- review \& editing.
\textbf{Noor Fatima:} Conceptualization, Data curation,
Investigation, Validation, Writing -- original draft, Writing --
review \& editing.
\textbf{Muhammad Bilal:} Conceptualization, Investigation,
Methodology, Validation, Writing -- original draft, Writing --
review \& editing.

\section*{Data and code availability}

The tool, the corpus manifest, the labelled pairs, and the full
analysis pipeline are available at
\url{https://github.com/amughalbscs16/cukereuse-release}. The corpus
parquet files (approximately 46~MB combined) are hosted in the same
repository under \texttt{corpus/}. The verbatim raw
\texttt{.feature}-file bodies (approximately 418~MB of directly
copyleft-inheriting content) are not redistributed; they are
reconstructed on demand by the \texttt{rehydrate.py} script from
the pinned commit SHAs recorded in the parquet, fetched from
GitHub's raw-content endpoint at each file's original repository.

\section*{About the authors}

\paragraph{Ali Hassaan Mughal} is an independent researcher currently
pursuing an Applied MBA in Data Analytics at Texas Wesleyan
University, USA. He earned his M.Sc. in Computer Science from Kansas
State University. Prior to his current research endeavours, he
served as a Senior Software Developer and Team Lead at Xpressdocs
and as a Software Developer II at Paycom. His research interests
include automated software testing, web application quality
assurance, and applied machine learning.

\paragraph{Noor Fatima} is an undergraduate student pursuing a
B.E. in Computer Engineering at the National University of Sciences
and Technology (NUST), Pakistan. She has a strong interest in
software development and hardware systems, with a growing focus on
emerging technologies in computing.

\paragraph{Muhammad Bilal} is an independent researcher pursuing an
M.Sc. in Management at the Technical University of Munich, Germany.
He has professional experience as both a Business Analyst and a
Software Engineer. His research interests focus on the impact of
technology on business performance, product quality analytics, and
the automation of industrial pipelines and processes.

\end{document}